# Using LDA and LSTM Models to Study People's Opinions and Critical Groups Towards Congestion Pricing in New York City through 2007 to 2019

**Ye Q.,[1] Chen X. H., Ph.D.,[2*] Kalan O., Ph.D.[3] and Ozbay K., Ph.D.[4]**

[1]Key Laboratory of Road and Traffic Engineering of the Ministry of Education, Tongji University, 4800 Cao'an Road, Shanghai 201804, China; email: yeqian122@163.com
[2*]Key Laboratory of Road and Traffic Engineering of the Ministry of Education, Tongji University, 4800 Cao'an Road, Shanghai 201804, China; email: tongjicxh@163.com
[3]C2SMART (A Tier 1 USDOT UTC), the Department of Civil and Urban Engineering, New York University, Brooklyn, 11201 NY, USA; email: ok22@nyu.edu
[4]C2SMART (A Tier 1 USDOT UTC), the Department of Civil and Urban Engineering, New York University, Brooklyn, 11201 NY, USA; email: kaan.ozbay@nyu.edu

## ABSTRACT

This study explores how people's views and responds to the proposals of NYC congestion pricing evolve over time. To understand these responses, Twitter data is collected and analyzed. Critical groups in the recurrent process are detected by statistically analyzing the active users and the most mentioned accounts, and the trends of people's attitudes and concerns over the years are identified with text mining and hybrid Nature Language Processing techniques, including LDA topic modeling and LSTM sentiment classification. The result shows that multiple interest groups were involved and played crucial roles during the proposal, especially Mayor and Governor, MTA, and outer-borough representatives. The public shifted the concern of focus from the plan details to a wider city's sustainability and fairness. Furthermore, the plan's approval relies on several elements, the joint agreement reached in political process, strong motivation in real-world, the scheme based on balancing multiple interests, and groups' awareness of tolling's benefits and necessity.

## INTRODUCTION

A congestion pricing proposal in New York City (NYC) was ultimately passed in April 2019 and the plan will be implemented as soon as January 2021 (Fix NYC 2018). However, it encountered public and political problems before its approval. The proposal has been revised for 4 rounds over the past 13 years. As the first American city approved congestion pricing, it is significant to explore how this proposal goes through the policy process and what are key factors to the results.

Historically, congestion pricing was introduced in 1975, successful practices worldwide includes Singapore's Area Licensing Scheme and Electronic Road Pricing, London Congestion Charge, Stockholm Congestion Tax, Milan EcoPass and Area C, Gothenburg Congestion Tax, and the Norwegian cities' urban tolls (Lehe 2019). Recently, several cities have already gained the approval of plans and some will be implemented soon, like in France, Jakarta, England, Vancouver, Abu Dhabi, etc. In the transport policy domain, there are three lessons learned from both successful and not implemented practices; (1) Both public and political acceptability are key to successful implementation (Hensher and Bliemer 2014), (2) Not all of the pricing proposal can be implemented effectively, depending on the complex political process, impacts of cultural diversity, dominate effects coming from case-specific factors, etc. (Liu et al. 2019; Noordegraaf et al. 2014), (3) Policy implementation will be successful only when many factors positively contribute to the process (Noordegraaf et al. 2014).

There are also several studies interested in the NYC case; interestingly, the findings are quite different from the previous ones in other cities. It is supposed that NYC's congestion pricing will be an important empirical case to complement the existing studies. For example, Winslott-Hiselius et al. (2009) demonstrated that New Yorkers living in the charged area were more acceptable to the plan than those outside the area. Schaller (2010) analyzed the event that the Bloomberg's 2007 proposal gained widespread public support but was ultimately blocked in the State, and conducted that even small groups of auto users can block the plan through the political process.

With the limited dataset and techniques, most of the previous literature conducted short-term quantitative analysis and case studies based on public acceptance surveys (Eliasson and Jonsson 2011), interviews (Liu et al. 2019) or first-hand observations (Schalle 2010). Some data-driven methods were utilized, including regression models (Schuitema and Rothengatter 2010), Structure Equation Model (Kim et al. 2013), discrete choice models (Di Ciommo et al. 2013), etc. Within the past few years, Mobile Social Networks (i.e., Twitter, Facebook), with a large number of active users, allow people to communicate with various groups and follow the trending topics. Meanwhile, these social networks can be used to obtain feedback from people about policy implementation by governments (Syaifudin and Puspitasari 2017). Opinion mining and sentiment analysis using social media became a field of interest for many researchers (Pak and Paroubek 2010; Ye et al. 2019).

Based on the policy lessons and unique findings in NYC case, the objective of this paper is to uncover critical factors of congestion pricing in NYC and further to verify the previous lessons focusing on the policy process. It aims to answer three questions, (1) to identify how New Yorkers perceive and respond to the plan during the recurrent process; (2) to explore the critical groups who have an effect or being affecting by the proposals; (3) to discover people's concerns and the switch of concerns. To achieve this goal, we collect Twitter data which contains the keywords of congestion pricing from January 2007 to November 2019, and then applied a integrated framework including text mining, LDA and LSTM techniques to conduct analysis to answer the aforementioned questions. This study can help explore potential interest appeal of the public, detect key factors and social influence in the policy process.

## BACKGROUND

New York City congestion pricing efforts including bridge tolls and parking bans can date to the early 20th century. This paper summarizes the timeline since 2007 consisting of 4-round of plan revisions and called it the 'policy process'.

(1) Bloomberg's 2008 proposal. NYC Mayor Michael Bloomberg firstly proposed a scheme in April 2007 by as part of a city's sustainability plan, entitled PlaNYC 2030. Vehicles entering Manhattan south of 86$^{th}$ street were charged once per day at 4-8 dollars. Taxis, For Hire Vehicles (FHVs), and vehicles passing East River bridges were exempted. The proposal gained widespread public support as well as passed voting of the city council; however, it didn't success. Because small groups of outer-borough Assembly Democrats who are auto users blocked the political process, the plan was never put to a vote in the New York State Assembly.

(2) 2015 Move NY plan. A former traffic commissioner Sam Schwarz suggested tolling all East River bridges. The plan charged the charging zone to the area below 60th street primary and then between 60th and 96th streets. Commercial vehicles paid daily rate and private vehicles paid per use at 5.54-8 dollars toll. The 75% of the estimated revenue would go to the MTA and 25% to the city DOT. Only yellow taxis are exempted. However, the borough president of Queens opposed it. Although the bills to implement Move NY were introduced at the state level in 2016, the proposal was failed in the state legislature.

(3) 2017 proposal. In response to the transit emergency (i.e., delays and crowded problems) for the MTA (Metropolitan Transportation Authority) burst in summer 2017, Governor Andrew Cuomo drafted a congestion pricing scheme similar to Bloomberg’s. The purpose was to raise funds for city transit and reduce street gridlock while to balance commuting. Cars, trucks and taxis paid tolls or surcharge per trip. The public overall has generally supported Cuomo's plan. However, Mayor de Blasio (Bloomberg’s successor) and Queens politicians opposed it. Meanwhile, news media has mixed reviews: NY Times reported that Bloomberg's proposal would have prevented the transit crisis, Slate Magazine lauded the pricing while The New York Post doubted it.

In October 2017, the State Government created a FixNYC Advisory Panel to find solutions for transit and congestion. The panel included various groups of people, including Schwartz et al. who had supported the congestion pricing even after Bloomberg's plan had been defeated. The proposal would firstly charge taxis and FHVs, including TNCs (i.e., Uber, Lyft), below 96th street and then cars and trucks below 60th street (Fix NYC 2018).

(4) Current proposal. In January 2019, Cuomo announced another state budget. The following month, Cuomo and de Blasio jointly agreed on a plan that stepped in fixing MTA operations. In March, Cuomo's pricing plan was again included in the 2019 state budget. In April, the proposal was ultimately passed the state legislature. The Traffic Mobility Act in 2019 is expected to fund 15 billion dollars of the next MTA project. Vehicles would toll once per day. Low-income residents in the zone were exempted. The plan outlined alternatives of four pricing scenarios, crediting driver for tolls and user-based exemptions.

**DATA**

Public discussion content can be extracted from twitter data. This study collected all the tweets that talk about congestion pricing from 2007 to 2019 through Twitter’s public facing application programming interface (API). Related tweets are searched by keywords, including ‘congestion pricing’, ‘congestion charging’, ‘congestion charge’, and ‘road charging’. 226 thousand tweets published by 119 thousand users are collected. They are discussing congestion pricing cases all over the world, such as London, Singapore, and San Diego. In order to extract the tweets specifically involved in the NYC case, a simple non-spatial filtering method is used. The tweets that contain the terms of ‘NYC’, ‘Brooklyn’, ‘Manhattan’, etc. is extracted; besides, texts that only mention terms related to other practices, i.e., ‘TfL’ (Transport for London), are removed from the raw data. The extracted data with corresponding filters are shown in Table 1. There are 44 thousand tweets in the final dataset. These tweets are directly related to the NYC congestion pricing for the last 13 years.

**Table 1. Data filter processing summary.**

| | Num of tweets | Ratio | Num of users | Ratio |
|---|---|---|---|---|
| Raw Tweets | 225,944 | 100% | 119,545 | 100% |
| NYC keyword filtering | 47,731 | 21.13% | 24,578 | 20.56% |
| Removal of other cases | 44,181 | 19.55% | 22,344 | 18.69% |

In the preliminary data exploration, the number of tweets, categories, hashtags, mentions, and users are statistically evaluated. In Figure 1, Twitter users did not post updates much about the NYC case before summer 2017, and then there were apparent bursts on Aug. 2017 (NYC Subway's state of emergency), Apr. 2018 and Apr. 2019 (plan's approval). Furthermore, Figure 2 shows that there were prominent discussed topics (e.g., Subway, Uber) and critical participants (e.g., Governor, Mayor). The following methodology and analysis sections will further investigate the topics, involvers, and opinions to find latent mechanisms on how the public, the government, and the agencies participate in the policy formation.

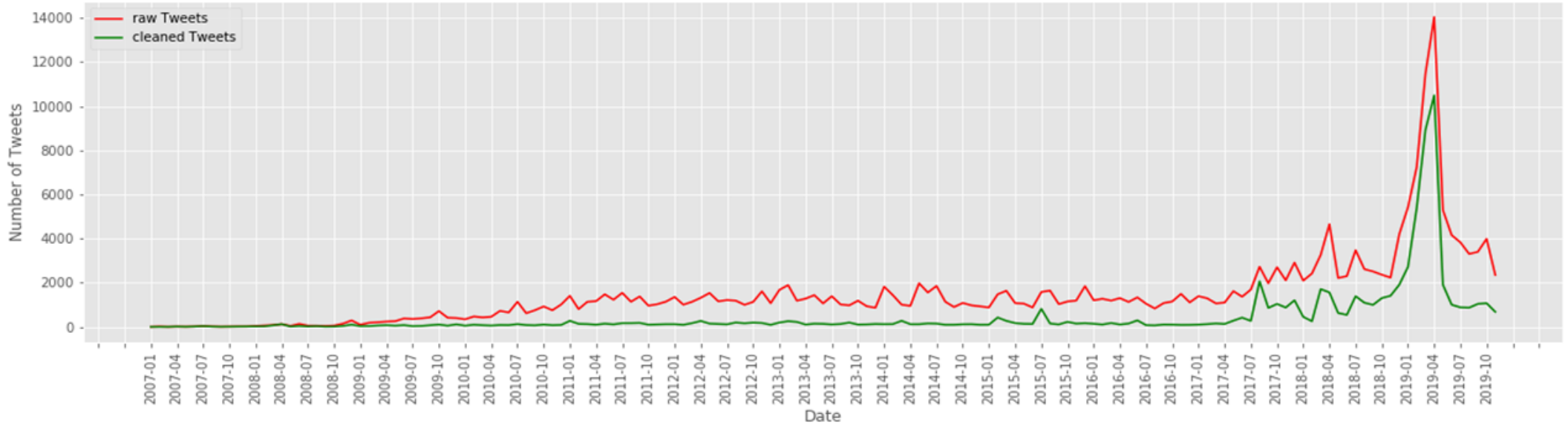

**Figure 1. Time series of number of cleaned data and raw data.**

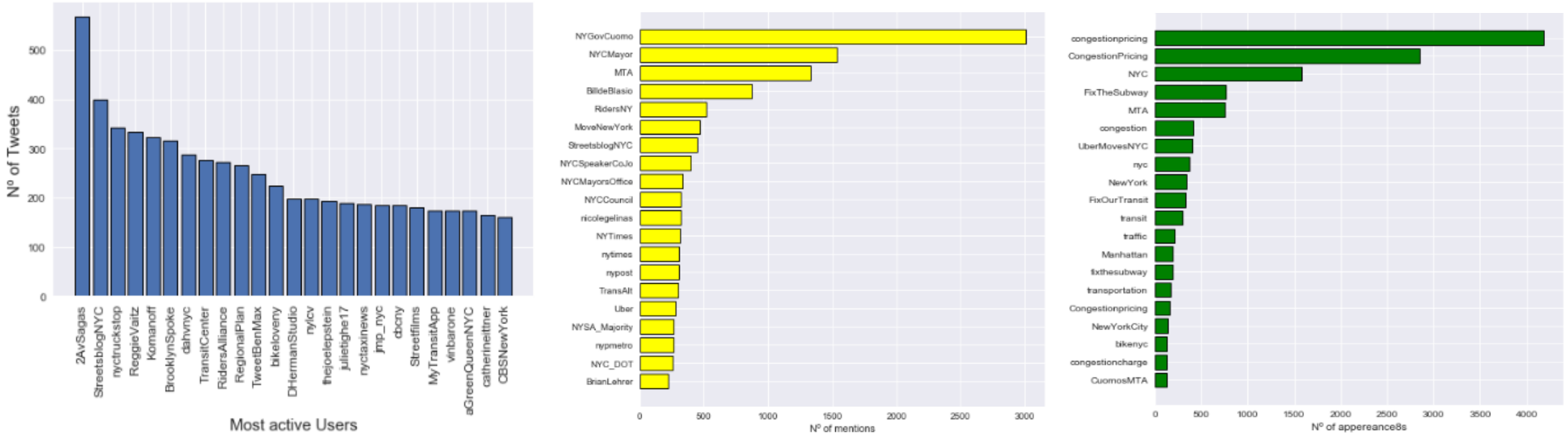

(a) Number of tweets for active users (b) Most mentioned accounts (c) Most used hashtags

**Figure 2. Statistical analysis of tweet categories, users, mentions, and hashtags.**

## METHODOLOGY

### Research framework

This study gradually digs into public concern and attitude variance with three main steps of the background study, text processing, and NLP analysis (Figure 3). In the previous section, the timeline is constructed to study the recurrent policy process among 2008, 2015, 2017, and 2019. The plan proposer, scheme details (i.e., pricing, charging area, exemptions), public acceptance, supporters and opponents, results of the plan and dominant reasons are summarized. After that Twitter data is collected, cleaned, and pre-processed for modeling. Publishers and mentioned users are analyzed to explore the critical groups those are affecting or affected by the proposals, LDA topic model is built to discover the changes of the most emphasized concerns, and LSTM model is built to detect individuals and groups' concern of focus through the years.

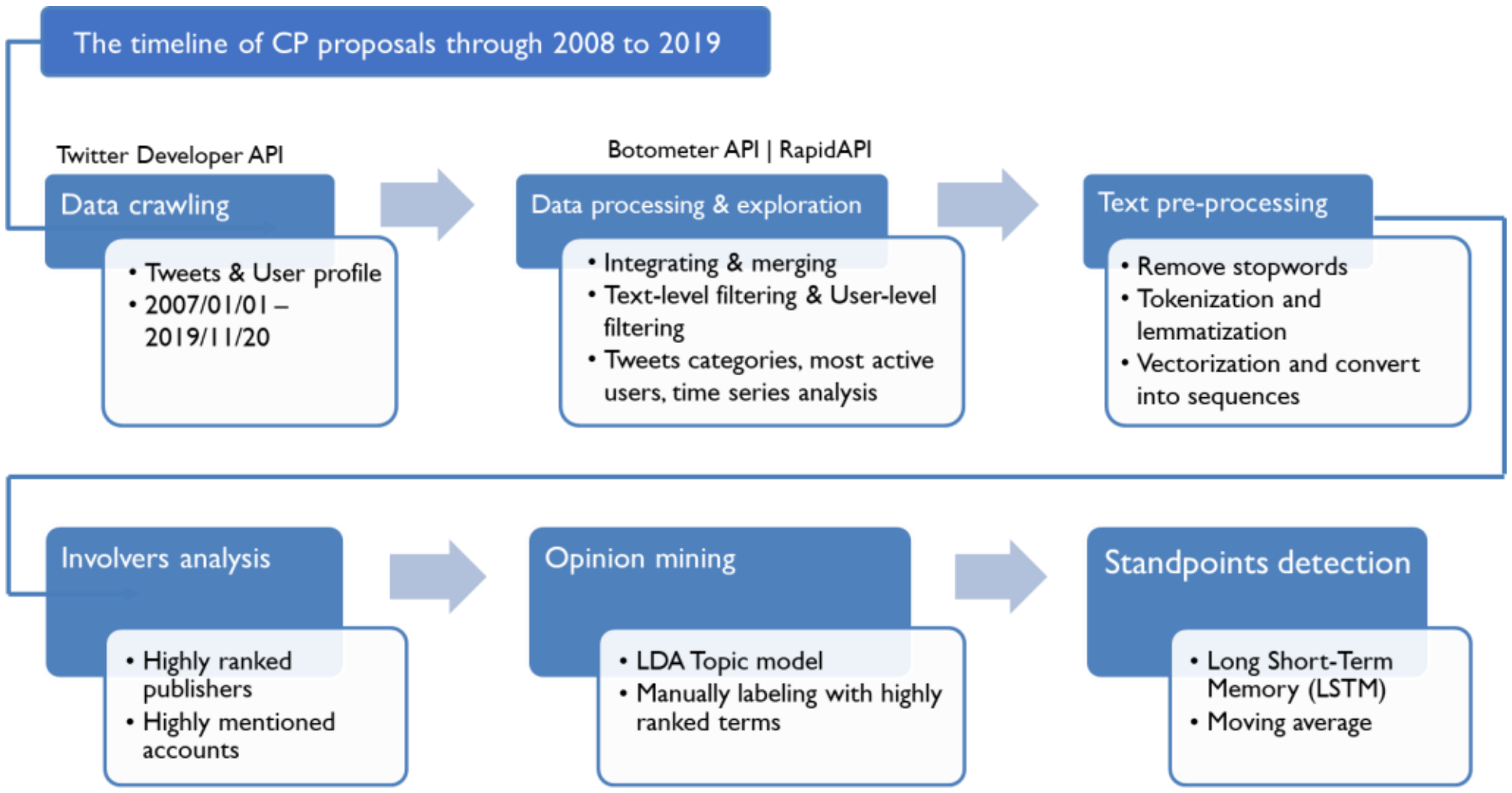

**Figure 3. Overall research framework.**

**Topic clustering using LDA model**

Latent Dirichlet Allocation (LDA), introduced by Blei et al. (2003), is an unsupervised learned model that can identify topics in documents using 'bag of words' method. It has advantages to output terms to describe a topic, make soft-clustering instead of hard-clustering that each entity can only belong to one cluster.

For the modeling process shown below, assuming the number of topics among all the document is $K$, distribute these $K$ topics across document $m$ by assigning each word a topic. For each word $n$ in document $m$ (per-document topic distribution $\alpha$), assume its topic is wrong but every other word is assigned the correct topic. Probabilistically assign a specific word $w$ a topic-based what topics are in document $m$ and how many times word $w$ has been assigned a particular topic across all of the documents (per-topic word distribution $\eta$). Iterate this process for each document until converging. The topic coherence is used to justify the quality of topics generated by the LDA model, UMass measure (Stevens 2012) based on document co-occurrence is choose, seen Equation 1-2.

$$coherence(V) = \sum_{(v_i,v_j)\in V} score(v_i, v_j, \epsilon) \quad (1)$$

$$score\left(v_i, v_j, \epsilon\right) = log\frac{D(v_i,v_j)+\epsilon}{D(v_j)} \quad (2)$$

where $V$ is a set of word describing the topic and $\epsilon$ indicates a smoothing factor which guarantees that score returns real numbers; $D(x, y)$ counts the number of documents containing words $x$ and $y$ and $D(x)$ counts the number of documents containing $x$.

A default model was trained using Gensim (Rehurek and Sojka 2010) implementation with default hyperparameters $\alpha$ =0.01 and $\eta$=0.1 under $K$=10, the baseline coherence score is 0.35. To improve the model, a series sensitivity tests was performed to help determine the model hyperparameters, one parameter (i.e., $K$) at a time by keeping others constant and run them ($\alpha$ and $\eta$) over the validation set. This study used grid search with validation set of 20% corpus (10,000 samples). Figure 4 outlines the coherence score in default model, $K$=9 with the highest score of 0.38 was picked as the optimal number of topics. Next, we selected the values of α and $\eta$ that yield maximum coherence for $K$=9. As a result, the hyperparameter combination is $\alpha$ ='asymmetric', $\eta$ =0.81, and $K$=9, which yields approximately 30.9% improvement over the baseline score.

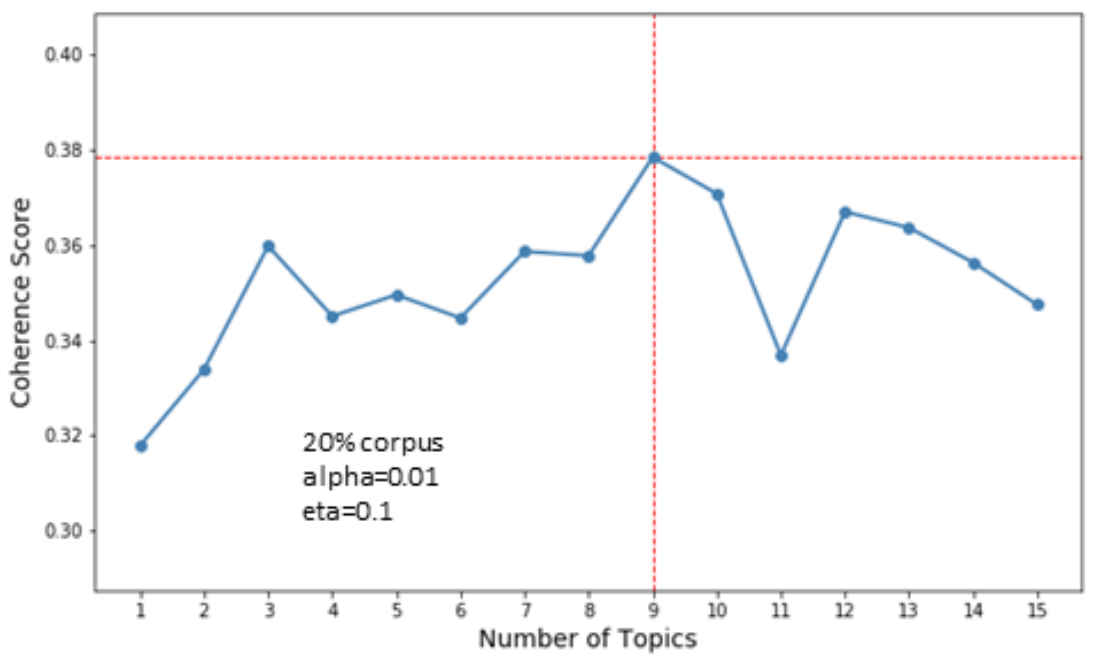

**Figure 4. Coherence over K=9.**

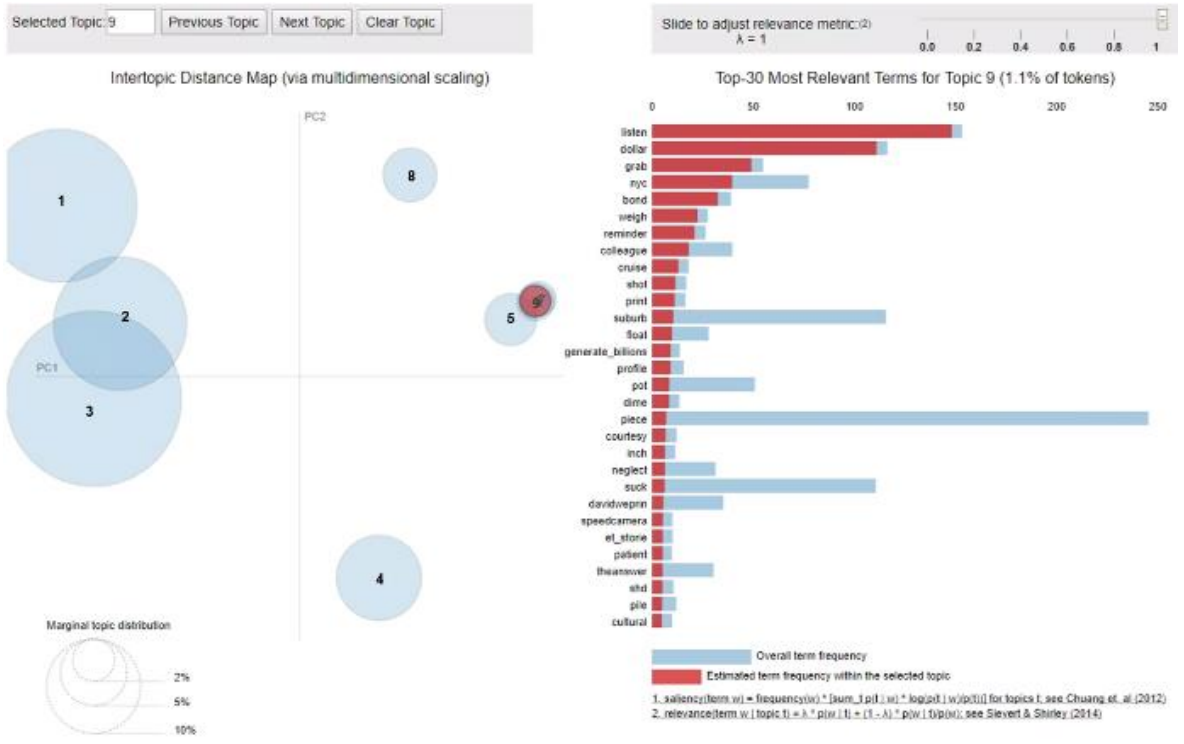

**Figure 5. Inter-topic distance map byLDAvis.**

To display the relationship between topics based on the training result under K = 9, a python-based topic model visualization package pyLDAVis (Sievert and Shirley 2014) was used. The radius of a circle represents the quantity of documents under a certain topic. The inter-topic distance indicates the similarity between topics, which is computed by default Jensen-Shannon divergence. Figure 5 shows that most circles are reasonably big, clearly separated and far from each other. In other word, the Twitter data regarding congestion pricing has significant topic differentiation and the trained LDA model works well to cluster the topics.

**Sentiment analysis using LSTM model**

Sentiment analysis, also known as opinion mining, is crucial to understanding user generated text in social networks by inferring the sentiment polarity (e.g. positive, negative, neutral). Researchers applied many approaches consisting of lexicon-based approach, Naïve Bayes approach, maximum entropy, Support Vector Machines, etc (Ahuta et al. 2014). In this study, we applied Long Short-Term Memory (LSTM) to infer the public's sentiment towards congestion pricing proposal as it is a state-of-the-art performer for semantic composition in the field of Twitter sentiment classification (Tang et al., 2015). It is capable of processing the entire sequences of data (e.g., text data) and predicting sentiment based on semantics rather than words.

LSTM is a recurrent neural network architecture proposed by Hochreiter and Schmidhuber (1997). A general LSTM architecture is composed of a *memory cell* and three gates, including an *input gate*, *output gate*, and *forget gate*. The *memory cell* is responsible for keeping track of the dependencies between the elements in the input sequence. The *input gate* controls the extent to which a new value flows into the cell, the *forget gate* controls the extent to which a value remains in the cell, and the *output gate* controls the extent to which the value in the cell is used to compute the output activation of the LSTM unit. The key to LSTMs is the cell state *C*, the horizontal line running through the top of the *memory cell*. There are connections into and out of the LSTM gates, a few of which are recurrent. The learned weights of these connections determine how the gates operate. Figure 6 below shows the common internal cell structure.

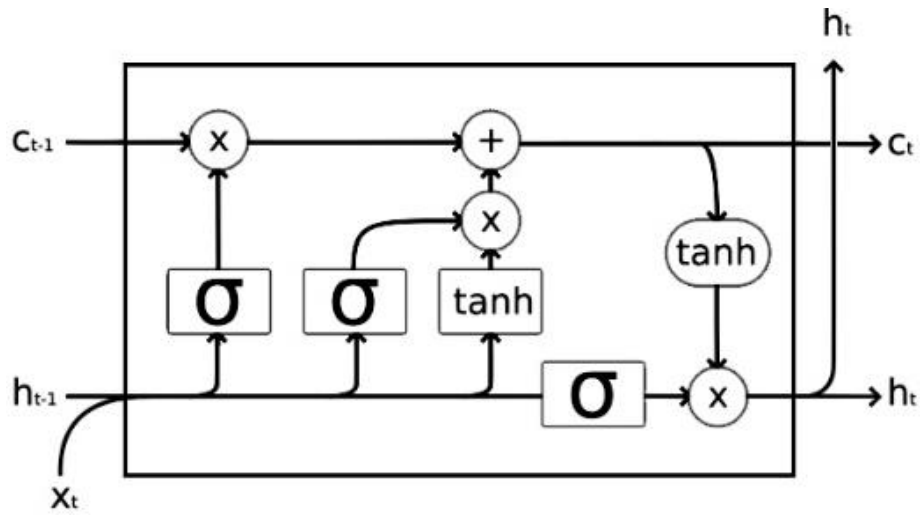

**Figure 6. LSTM cell structure.**

Where *C* is the cell state vector, *h* is the hidden state vector (output vector), *x* is the input vector, $\sigma$ is the activation function (i.e., sigmoid function shown in Equation 3), *tanh* is the tanh layer used to process the summed weighted input.

$$S(x) = \frac{1}{1+e^{-x}} \quad (3)$$

The LSTM model was trained with the Twitter Sentiment Analysis Corpus manually labeled by Sanders (2011) which contains 1,578 thousand classified tweets marked with identified sentiment of 1 for positive and 0 for negative. The corpus was split into train sets (2/3) and test sets (1/3). To output the prediction probabilities, the softmax activation function was used. After hyperparameter tuning, the model took parameters of the max feature of 2000, batch size of 32, iteration epoch of 7. The accuracy of test data is 0.81. Next, to evaluate the prediction performance in our data set, the trained model was tested on our tweet data regarding congestion pricing. We manually labeled 124 rows of tweets with 76 of positive sentiment and 48 of negative sentiment. It achieved the accuracy of 0.78, precision of 0.83, recall of 0.79, and F1-score of 0.81. The model works well for the prediction in NYC case. Finally, the sentiments of all 44 thousand tweets were identified polarities as 0 or 1.

# ANALYSIS RESULTS

## Involvers analysis

We assumed that the important Twitter involvers in the policy process of congestion pricing are the users either who tweeted about congestion pricing at many times (i.e., active users) or which accounts were mentioned frequently in the tweets regarding congestion pricing (i.e., mentioned accounts). Therefore, top-10 active users and top-10 mentioned accounts in each year were extracted, and their involving percentages were calculated by the number of publishing tweets and mentioning tweets, and finally were summed up through the years (Figure 7). Most active users represented the people whose interests would be possibly affected by congestion pricing, so that they post tweets frequently to show supports or negative voice. Meanwhile, the most mentioned accounts mean those who are believed to play key roles (i.e., proposers, government officials, news media) in the policy formation, the public want them to speak out.

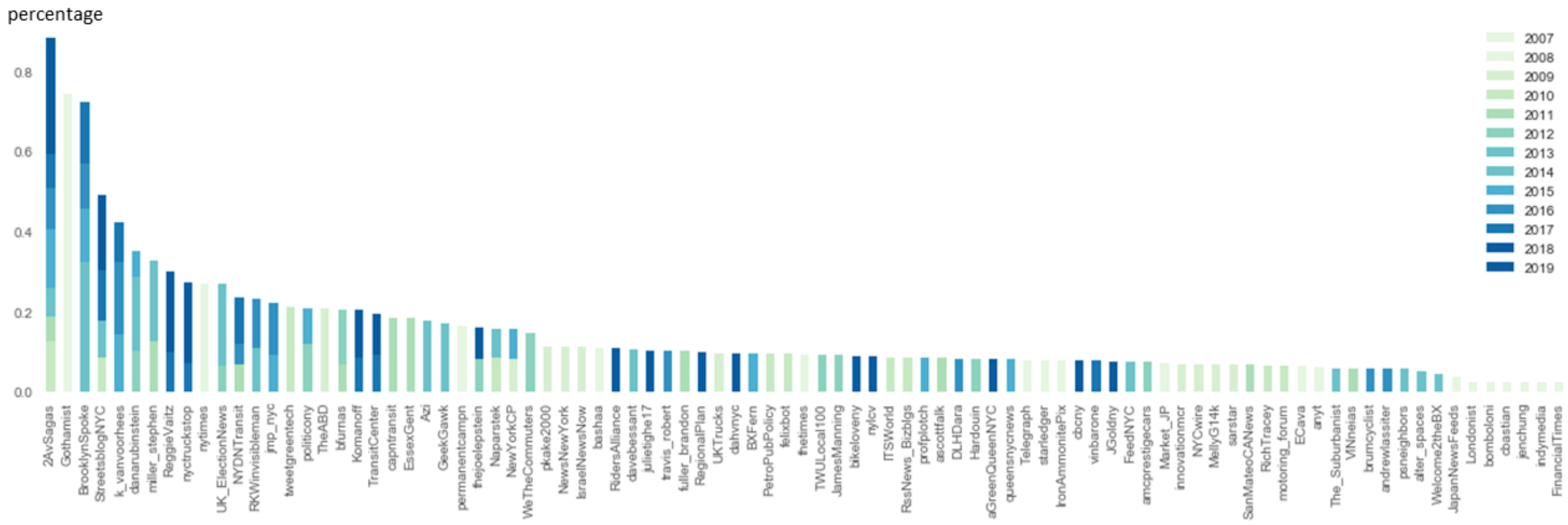

(a) top-10 active users

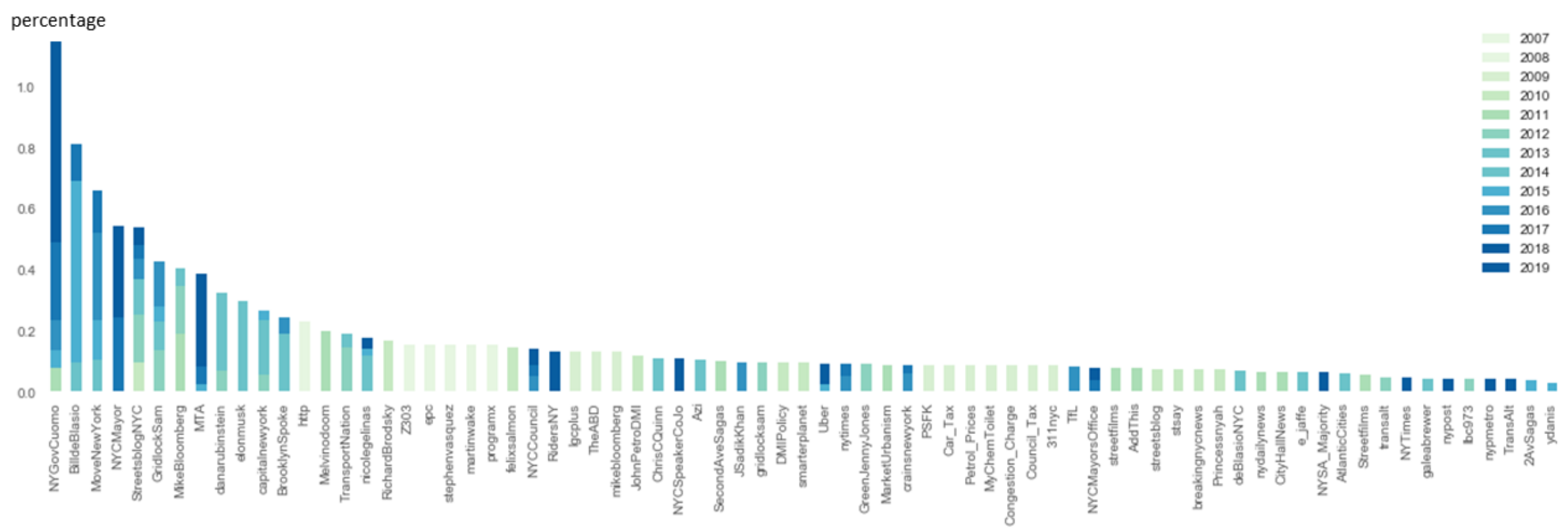

(b) top-10 mentioned accounts

**Figure 7. The most important involvers who tweet about congestion pricing.**

In Figure 7(a), before 2010, news sources were the most interested groups, like *Gothamist*, *The New York Times* (NY Times), *New York Daily News*, followed by *Mr. Furnas* (policy & planning official for NYC Mayor) and *Jon Sorensen* (ITS Engineer). Since Bloomberg's 2008 proposal was the first congestion pricing plan in North American cities, many regional and local news media followed up actively. At the middle phase of policy process, *Second Ave. Sagas* (official mayoral source), *Mr. Gordon* (TV writer/producer), *Streetsblog New York City* (Streetsblog NYC), *Ms. Voorhees* (urban designer), and *NYDN Transit* (Transit news from NY Daily News) showed much more interests. When the proposal experienced the third revision, *Second Ave. Sagas* and *Streetsblog NYC* kept updating on the progress to the public. Besides, individuals or communities who tweets mainly about taxi and FHVs (*Ms. Vaitz*), trucks and commercial vehicles (*New York Truck Stop*), public transit (*TransitCenter*) and *Mr. Komanoff* who is a policy analyst and carbon-taxer were identified as active users.

One of the crucial involvers reflected on Twitter in these years was the Mayor de Blasio as he firstly opposed Cuomo's 2017 plan and called a wealth tax instead, and then shifted to agreement after the joint negotiation. Combined with the background section, once the NYC's tolling implemented, drivers of taxis and FHVs, truck and vehicles would have to pay congestion fees while public transit system would be funded. As a result, these two groups being affected differently expressed opinions many times on Twitter.

Figure 7(b) shows that many accounts were mentioned during the first of plan's first revision while only several accounts were significantly addressed later. Early mentions were of various backgrounds, including the officials of city and state (e.g. *Governor Cuomo*, *Mayor Bloomberg*, Former NYS Assemblyman *Mr. Brodsky*), news sources (e.g. *Streetsblog NYC*), other pricing policies such as the car tax and petrol prices, and official government services *NYC 311*. Although the 2008 proposal gained widespread acceptance as well as the great promoting by Cuomo and Bloomberg, a part of users wanted to communicate with the Assemblyman who had expressed doubt of congestion pricing in terms of the financial fairness and environmental impact. When the plan was revised for the second and third time, *Cuomo* and *de Blasio* were the most frequently mentioned accounts as they held different opinions on pricing implementation and have negotiated for nearly two years; followed by the supportive accounts *Sam Schwarz* and his plan *Move NY* who probably pushed the plan. Besides, news media (e.g. *Streetsblog NYC*, senior reporter *Ms. Rubinstein*) were also addressed to update events.

When coming to the fourth proposal, two crucial people, Cuomo and de Blasio, as well as MTA occurred the most. Besides, *Uber* was also mentioned many times as it would be charged at the second phase of plan (FixNYC 2018). The key to the ultimate approval of the plan resulted from not only the joint agreement reached in the political process; more importantly, MTA's impetus due to the massive financial shortage of 15 billion dollars.

**Opinion mining**

The trained LDA model categories 9 dominant topics and presents the most related keywords among the tweets, and then, topics are manually referred to a label (Table 2). The most prevalent topics are: Topic 1 - Key elements of the proposals, such as setting a proper plan goal, achieving both the city-level and the state-level agreement, gaining approval or not; Topic 2 - Implementation and expected effects; for example, users tweeted that carrying out congestion pricing may be a solution to current road traffic or subway system, however, the plan also faces issues that never met before; Topic 3 - Plan details, i.e., to charge vehicles traveling within the congested areas. As shown in Figure 5, Topic 6, 7 and 9 are similar, they are public policies and individual travel alternatives in NYC, the impacts on the environment, human health and society, general dialogs on Twitter, respectively.

**Table 2. The LDA model generated topics with K = 9.**

| Topic | Top-10 Keywords | Manually Labels |
|---|---|---|
| 1 | congestion, pricing, plan, say, tax, state, price, city, new, pass | Key elements of the proposal |
| 2 | congestion, pricing, need, transportation, come, implement, traffic, city, push, may | Implementation & effects |
| 3 | charge, congestion, road, car, pricing, pay, people, drive, go, many | Plan content |
| 4 | transit, subway, fix, fund, system, funding, need, mta, public, revenue | system needs funds |
| 5 | charge, road, ny, open, phone, save, key, technology, course, charging | Plan technologies |
| 6 | priority, promote, endorse, actual, condition, e-bike, gridlocksam, medium, scooters, movenewyork | Public policies & mobility alternatives |
| 7 | member, environment, however, total, average, game, kid, child, annual, several | Sustainable environment & human health & society |
| 8 | pricing, congestion, pictwittercom, support, traffic, plan, progressive, commute, ease, regressive | Both-side standpoints |
| 9 | listen, dollar, grab, nyc, bond, weigh, reminder, colleague, cruise, shot | Dialogs with replies |

Figure 8 displays the popularity of each topic evolving over the years. The class of color is generated by quantile. A grid with darker color indicates that more tweets were posted under a certain topic in a specific year; for example, there are less than 20 tweets under topic 2 in 2007.

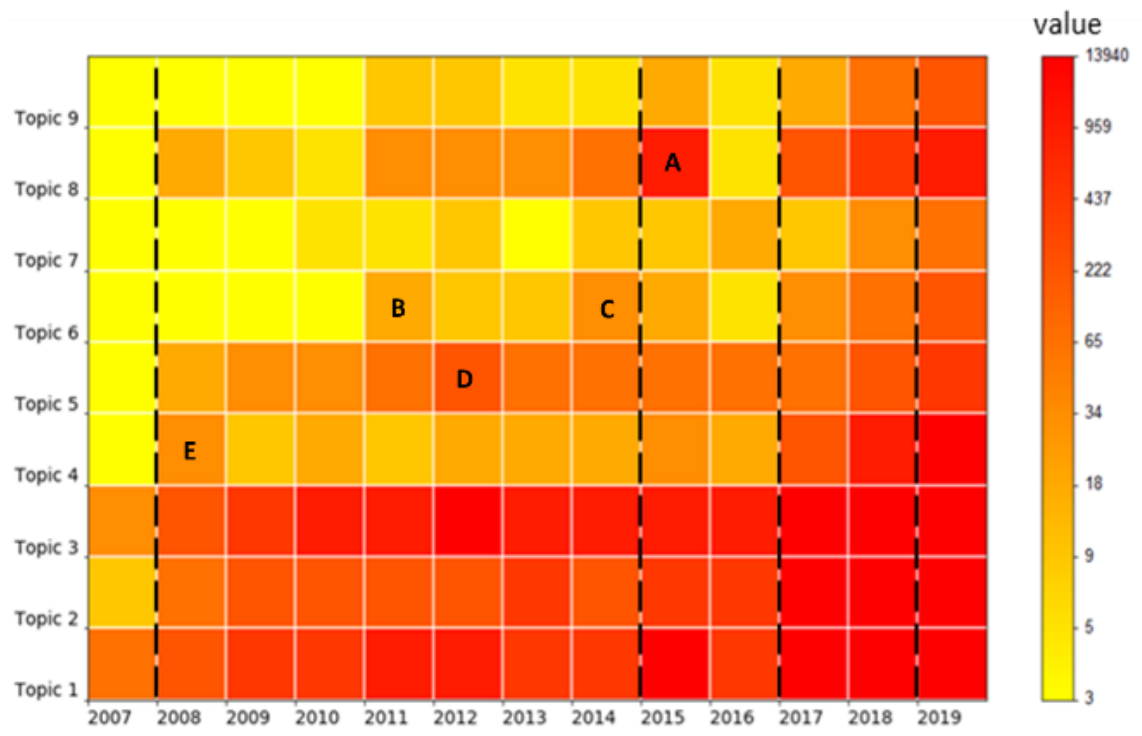

**Figure 8. Grid heatmap of 9 topics over the years.**

The reasonable trend is that a certain topic would receive more and more attention over the years due to the climbing number of Twitter users; for example, three most popular topics of Topic 1, 2 and 3. However, several topics become heat suddenly in several years and then heat down soon, e.g. the usual phenomenon in 2015 for Topic 8 ('A'), in 2011 ('B') and 2014 ('C') for Topic 6, in 2012 ('D') for Topic 5, in 2008 ('E') for Topic 4. Several findings can be summarized as follows. Firstly, since the Bloomberg's 2008 proposal released (grid 'E'), New Yorkers have concerned about fixing old transit and subway system and kept discussing till now. The urban problem of an urgent funding need is the case-specific factor of NYC's congestion pricing implementation. Secondly, grid 'A' reveals an intense debate on Move NY plan. One of the related events is that the East River bridges tolling was decried by Queens lawmakers at the end of 2015, because it damaged the interest of out-borough residents travelling to Manhattan. The setting of proposal is quite complex due to multiple interest groups involved. Thirdly, grid 'D' refers to the key techniques that enable congestion pricing in NYC. The previous proposals planed a E-ZPass transponders and a license plate recognition system; in Cuomo's scheme, more alternatives such as open road tolling technology, roadside Bluetooth readers, connected vehicle technology, smartphone applications would be took advantage.

**Sentiment analysis**

Sentiment analysis on people's feedback against public policy helps leaders understand public concerns and needs, as well as detect special events, so that to better implement urban policy. The LSTM model predicts the tweets of sentiment including 41,046 positives and 28,140 negatives. Figure 9 shows the time series of sentiment polarities over the years. The horizontal dash line shows a sentiment threshold of half positive and half negative, the vertical dash line divides the process into 4 rounds.

Overall, Twitter users hold generally more positive reactions to NYC congestion pricing, and it fluctuates randomly. A finding can be demonstrated is that when each round of proposal released, people's response within the following months on Twitter is supportive. Especially, the tweets published in October 2007, Summer 2008, November 2015, summer 2016 and summer 2018 reach a peak of positive sentiment. On the contrary, there are two special months, July 2015 and February 2017, when the positive curve drops rapidly below 40% and then climb to a rather high level, at around 70%. Some special events (i.e., public debates, voting) happened might have significant effect on New Yorkers during these months.

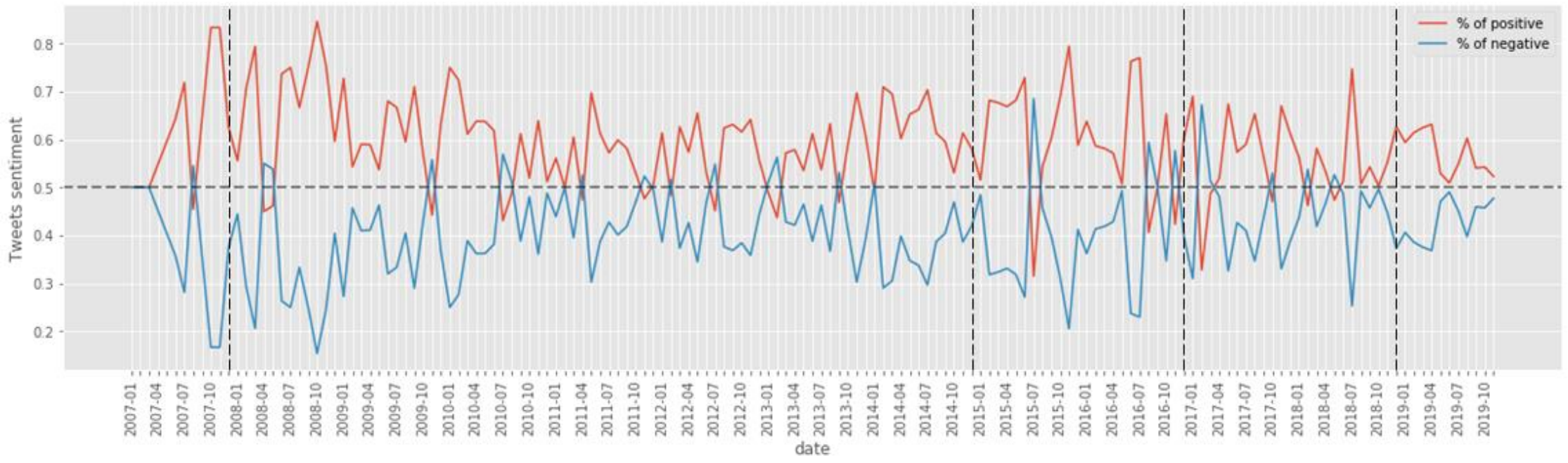

**Figure 9. Time series of sentiment polarities over years.**

## DISCUSSION AND CONCLUSION

Throughout the 13-year analysis of congestion pricing tweets, the findings can be drawn to answer three proposed questions in the NYC case. (1) The public's concern of focus have shifted from the proposal details into more comprehensive issues, such as funding for a reliable transportation system, city's sustainability and social fairness, effective public policies, new mobility alternatives. When the plan formed and just released soon, people would inevitably argue its content and reasonableness. Various interest groups expressed opinions and even debated intensively. With the gradual stabilization of the plan, attention would be paid more on the effects of policies and how to guarantee of implementation in a long run. (2) The leading roles are not only played by plan proposers; multiple parts of interest groups or individuals are significantly involved in the policy process. These groups in NYC case are government officials (e.g. Mayor and Governor, Assemblymen, out-borough representatives), transport associations (e.g. MTA, Gridlock Sam, Uber) and news media (e.g. Streetsblog NYC, NY Times). (3) Different from other practices (i.e., Stockholm congestion charging), New Yorkers' attitude toward congestion pricing are generally positive no matter whether the proposal was passed or not, how the plan was revised time after time.

When formulating a public policy, it is bound to be a recurrent process of interests' negotiation and consensus-building. In the New York case, the proposals was gained public acceptance over the years; however, the crucial factors that contributing to the ultimate approval are four elements: (1) obtained the support of a wider segment; (2) motivated by a strong promotion, for example, MTA's massive financial needs in fixing system; (3) joint agreement in the political process between Mayor and Governor; and (4) interest groups' awareness of benefits or necessity coming from congestion pricing and no longer strongly oppose. In addition, the scheme of congestion pricing including the charging area and user-based pricing are also key factors, which needs to be built on the premise of balancing the interests of all parties (i.e., auto users, out-borough citizens, mobility providers like Uber and Lyft) with the best efforts. Once these elements are achieved, the congestion pricing plan would be not far from implementation.

## ACKNOWLEDGEMENT

The authors acknowledge support from the China Scholarship Council (No.201906260112) and the project of the Natural Science Foundation of China (No.71734004). It was also partially supported by C2SMART Tier 1 University Transportation Center at New York University.

## REFERENCES

Fix NYC. (2018). Fix NYC Advisory Panel Report January 2018.

Noordegraaf, D. V., Annema, J. A., & van Wee, B. (2014). Policy implementation lessons from six road pricing cases. *Transportation Research Part A: Policy and Practice*, *59*, 172-191.

Lehe, L. (2019). Downtown congestion pricing in practice. *Transportation Research Part C: Emerging Technologies*, *100*, 200-223.

Hensher, D. A., & Bliemer, M. C. (2014). What type of road pricing scheme might appeal to politicians? Viewpoints on the challenge in gaining the citizen and public servant vote by staging reform. *Transportation Research Part A: Policy and Practice*, 61, 227-237.

Winslott-Hiselius, L., Brundell-Freij, K., Vagland, Å., & Byström, C. (2009). The development of public attitudes towards the Stockholm congestion trial. *Transportation Research Part A: Policy and Practice*, *43*(3), 269-282.

Schaller, B. (2010). New York City's congestion pricing experience and implications for road pricing acceptance in the United States. *Transport Policy*, *17*(4), 266-273.

Eliasson, J., & Jonsson, L. (2011). The unexpected "yes": Explanatory factors behind the positive attitudes to congestion charges in Stockholm. *Transport Policy*, *18*(4), 636-647.

Liu, Q., Lucas, K., Marsden, G., & Liu, Y. (2019). Egalitarianism and public perception of social inequities: A case study of Beijing congestion charge. *Transport Policy*, *74*, 47-62.

Schuitema, G., Steg, L., & Rothengatter, J. A. (2010). The acceptability, personal outcome expectations, and expected effects of transport pricing policies. *Journal of Environmental Psychology*, *30*(4), 587-593.

Kim, J., Schmöcker, J. D., Fujii, S., & Noland, R. B. (2013). Attitudes towards road pricing and environmental taxation among US and UK students. *Transportation Research Part A: Policy and Practice*, *48*, 50-62.

Di Ciommo, F., Monzón, A., & Fernandez-Heredia, A. (2013). Improving the analysis of road pricing acceptability surveys by using hybrid models. *Transportation Research Part A: Policy and Practice*, *49*, 302-316.

Syaifudin, Y. W., & Puspitasari, D. (2017). Twitter data mining for sentiment analysis on peoples feedback against government public policy. *MATTER: International Journal of Science and Technology*, *3*(1).

Pak, A., & Paroubek, P. (2010). Twitter as a corpus for sentiment analysis and opinion mining. In *LREc* (Vol. 10, No. 2010, pp. 1320-1326).

Ye, Q., Chen, X., Zhang, H., Ozbay, K., & Zuo, F. (2019). Public Concerns and Response Pattern toward Shared Mobility Security using Social Media Data. In *2019 IEEE Intelligent Transportation Systems Conference (ITSC)*(pp. 619-624). IEEE.

Blei, D. M., Ng, A. Y., & Jordan, M. I. (2003). Latent dirichlet allocation. *Journal of machine Learning research*, *3*(Jan), 993-1022.

Sievert, C., & Shirley, K. (2014). LDAvis: A method for visualizing and interpreting topics. In *Proceedings of the workshop on interactive language learning, visualization, and interfaces* (pp. 63-70).

Hochreiter, S., & Schmidhuber, J. (1997). Long short-term memory. *Neural computation*, *9*(8), 1735-1780.

Tang, D., Qin, B., & Liu, T. (2015). Deep learning for sentiment analysis: successful approaches and future challenges. *Wiley Interdisciplinary Reviews: Data Mining and Knowledge Discovery*, *5*(6), 292-303.

Rehurek, R., & Sojka, P. (2010). Software framework for topic modelling with large corpora. In *Proceedings of the LREC 2010 Workshop on New Challenges for NLP Frameworks*.

Stevens, K., Kegelmeyer, P., Andrzejewski, D., & Buttler, D. (2012, July). Exploring topic coherence over many models and many topics. In *Proceedings of the 2012 Joint Conference on Empirical Methods in Natural Language Processing and Computational Natural Language Learning*(pp. 952-961). Association for Computational Linguistics.

Bhuta, S., Doshi, A., Doshi, U., & Narvekar, M. (2014, February). A review of techniques for sentiment analysis Of Twitter data. In *2014 International conference on issues and challenges in intelligent computing techniques (ICICT)* (pp. 583-591). IEEE.

Sanders, N. J. (2011). Sanders-twitter sentiment corpus. *Sanders Analytics LLC*, *242*.